\documentclass[useAMS, usenatbib]{mn2e}

\usepackage{deluxetable}
\usepackage{epsfig} 
\usepackage{subfigure}
\usepackage{comment}
\usepackage{graphicx}
\usepackage{epstopdf}
\usepackage{lscape}
\usepackage{amsmath}
\usepackage{amssymb}
\usepackage{color}

\newcommand{\apj}{ApJ}

\newcommand{\apjl}{ApJL}
\newcommand{\apjs}{ApJS}
\newcommand{\mnras}{MNRAS}

\newcommand{\nat}{Nature}

\def\avg#1{\langle#1\rangle}
\newcommand{\Lam}{\Lambda}

\newcommand{\Msun}{M_{\odot}}

\newcommand{\hinv}{h^{-1}}
\newcommand{\himpc}{\hinv{\rm\,Mpc}}

\newcommand{\beq}{\begin{eqnarray}}
\newcommand{\eeq}{\end{eqnarray}}

\newcommand{\Om}{\Omega_{\rm m}}
\newcommand{\Ol}{\Omega_{\Lam}}

\newcommand{\Mhalo}{M_{\rm halo}}

\title{Pairwise Velocities of Dark Matter Halos: a Test for the \boldmath{$\Lambda$} Cold Dark Matter Model using the Bullet Cluster}

\author[Thompson \& Nagamine]{Robert Thompson$^1$\thanks{Email: rthompson@physics.unlv.edu}, Kentaro Nagamine$^1$ \vspace{0.3cm}\\
$^1$ Department of Physics \& Astronomy, University of Nevada, Las Vegas, 4505 S. Maryland Pkwy, Las Vegas, NV, 89154-4002, USA \\
}

\begin{document}

\maketitle 


\begin{abstract}

The existence of a bullet cluster (such as 1E0657-56) poses a challenge to the concordance $\Lambda$ cold dark matter model.  Here we investigate the velocity distribution of dark matter halo pairs in large $N$-body simulations with differing box sizes ($250\himpc - 2h^{-1}$Gpc) and resolutions.  We examine various basic statistics such as the halo masses, pairwise halo velocities ($v_{12}$), collisional angles, and pair separation distances.  
We then compare our results to the initial conditions required to reproduce the observational properties of 1E0657-56 in non-cosmological hydrodynamical simulations. 

We find that the high velocity tail of the $v_{12}$ distribution extends to greater velocities as we increase the simulation box size. We also find that the number of high-$v_{12}$ pairs increases as we increase the particle count and resolution with a fixed box size, however, this increase is mostly due to lower mass halos which do not match the observed masses of 1E0657-56.  We find that the redshift evolution effect is not very strong for the $v_{12}$ distribution function between $z$=0.0 and $z$$\sim$0.5. 

We identify some pairs whose $v_{12}$ resemble the required initial conditions, however, even the best candidates have either wrong halo mass ratios, or too large separations.
Our simulations suggest that it is very difficult to 
produce such initial conditions at $z=0.0, 0.296,$ \& 0.489 in comoving volumes as large as (2\,$h^{-1}$Gpc)$^{3}$.
Based on the extrapolation of our cumulative $v_{12}$ function, we find that one needs a simulation with a comoving box size of (4.48\,$h^{-1}$\,Gpc)$^3$ and $2240^3$ DM particles in order to produce at least one pair of halos that resembles the required $v_{12}$ and observed masses of 1E0657-56.
From our simulated $v_{12}$ probability distribution function, we find that the probability of finding a halo pair with $v_{12}\geq 3000$\,km\,s$^{-1}$ and masses $\geq 10^{14} \Msun$ to be $2.76\times10^{-8}$ at $z$=0.489.  
We conclude that either 1E0657-56 is incompatible with the concordance $\Lam$CDM universe, or the initial conditions suggested by the non-cosmological simulations must be revised to give a lower value of $v_{12}$.

\end{abstract}


\section{Introduction}
\label{sec:intro}

It is widely believed that the structure formation in our Universe is largely driven by the gravity of dark matter. Therefore it is worthwhile to probe dark matter dynamics through measurements of galaxy peculiar velocities and constrain our cosmological model by comparing against numerical simulations. In fact there has been extensive work along these lines, recovering the local density field from the measured velocity field \citep{Bert89,Davis96,Willick96}. 
Unfortunately the observations of peculiar velocity fields contain large uncertainties, and accurate determination of the cosmological mass density parameter $\Om$ turned out to be difficult using this method. 

More recently, clusters of galaxies have been used to prove the existence of dark matter itself, thanks to accurate measurements of projected dark matter density using weak and strong lensing techniques. Some clusters show signs of a cluster-cluster merger, where the baryonic component and  collisionless dark matter show different spatial distributions, strongly supporting the existence of dark matter. Furthermore, using the shock features seen in the gas, one can infer the collision velocity of two galaxy clusters \citep{Clowe04,Clowe06,Bradac06}. These new observations have brought renewed interest to dark matter dynamics and using it to check the standard $\Lam$ cold dark matter ($\Lam$CDM) cosmological model \citep{Efsta90,Ostriker95}. 
  
In particular, the observations of the massive cluster of galaxies 1E0657-56 seem to suggest a much higher relative dark matter halo velocity than one would expect in the $\Lambda$CDM model.  This system includes a massive sub-cluster (the "bullet") with $M_{\rm{bullet}} \simeq 1.5 \times 10^{14} \Msun$ that has fallen through the parent cluster of $M_{\rm{parent}} \simeq 1.5 \times 10^{15} \Msun$ roughly $150$ million years ago, and is separated by $\simeq 0.72$\,Mpc on the sky at an observed redshift of $z$=0.296 \citep{Clowe04,Clowe06,Bradac06}.
The uniqueness of this system comes from the collision trajectory being almost perpendicular to our line of sight.  This provides an opportunity to better study the dynamics of large cluster collisions.  The Chandra observations revealed that the primary baryonic component had been stripped away in the collision and resided between the two clusters in the form of hot X-ray emitting gas \citep{Mark06}.  This provides strong evidence for the existence of dark matter (DM); As the two clusters passed through each other, the baryonic components interacted and slowed down due to ram pressure, while  
the dark matter component was allowed to move ahead of the gas since it only interacts through gravity without dissipation.
One can infer the velocity of the bow shock preceding the `bullet' using the shock Mach number and a measurement of the pre-shock temperature.  The inferred shock velocity was found to be $v_{\rm{shock}} =4740^{+710}_{-550}$\,km\,s$^{-1}$ \citep{Mark06}.  

\citet{Hayashi06} examined the Millennium Run \citep{Mill05} in search for such a sub-cluster moving with a velocity relative to its parent cluster of $v_{\rm{bullet}}=4500^{+1100}_{-800}$ km s$^{-1}$ \citep{Mark04}.  Due to the limited volume of the simulation (500 $h^{-1}$Mpc)$^{3}$, few halos had masses comparable to 1E0657-56.  Still they estimated that about 1 in 100 have velocities comparable to the bullet cluster, and concluded that the event is not impossible within the current $\Lambda$CDM model.

It is often assumed that the inferred shock velocity is equal to the velocity of the dark matter `bullet' itself.  Several groups have shown, however, that this is not necessarily true through the use of non-cosmological hydrodynamic simulations.  
\citet{Milo07} used two dimensional simulations to find that the subcluster's velocity differed from the shock velocity by about 16\%, bringing the relative velocity of DM halos down to $\sim$\,3980\,km\,s$^{-1}$.  They assumed a zero relative velocity at a separation distance of 4.6\,Mpc for their initial conditions.  They also emphasized that their conclusion is sensitive to the initial mass and gas density profile of the two clusters.
\citet{Springel07} was able to reproduce the inferred shock velocity through the use of an idealized three dimensional hydrodynamic simulation with initial conditions that assumed a relative velocity of 2057\,km\,s$^{-1}$ at a separation distance of 3.37\,Mpc, and found that the subcluster was moving with a relative speed of only $\sim$\,2600\,km\,s$^{-1}$ just after the collision. 
\citet{Mast08} argued that \citet{Springel07} failed to reproduce the observed displacement of X-ray peaks that represent an important indicator of the nature of the interaction.  In their simulations they found that in order to reproduce the observational data of 1E0657-56 a relative halo infall velocity of $\sim$\,3000\,km\,s$^{-1}$ at an initial separation distance of 5\,Mpc was required.

Similar to previous work by \citet{Hayashi06}, \citet{Lee10} quantified the likelihood of finding bullet-like systems in the large cosmological N-body simulation MICE \citep{MICE10}.  They examined DM halos at $z$=0.0 \& 0.5, searching for a halo pair matching the initial conditions of \citet{Mast08}.  They concluded that $\Lambda$CDM is excluded by more than 99.91\% confidence level at $z$=0.  Their results at $z$=0.5 are inconclusive due to limited statistics.  However, by fitting their pairwise velocity probability distribution function to a Gaussian distribution, they were able to estimate the probability of finding a pair with $v_{12}>3000$ km s$^{-1}$ to be $3.6\times10^{-9}$ and $v_{12}>2000$ km s$^{-1}$ to be $2.2\times10^{-3}$ at $z$=0.5.  They did warn that one must be careful about this approach since they are probing the tail of the distribution where their fits may not be accurate.

Most recently \citet{Forero10} approached the problem from a different perspective. 
They studied data from the MareNostrum Universe \citep{MareNostrum} which contains baryonic matter in addition to collisionless DM.  
Instead of examining the pairwise velocities of DM halo pairs, they concerned themselves with the physical separation between the dominant gas clump and its predominant DM structure.
They argued that their approach provides a more robust comparison to observation; deriving the relative velocity from the observations includes statistical and systematic uncertainties whereas the separation uncertainty is dominated by statistical errors in the measuring process.
Additionally they point out that current simulations do not include the proper prescriptions for cooling, star formation, or feedback which implies that their predictions of the detailed X-ray properties of hot gas in massive halos are not robust.
Using their method they found that large displacements between gas \& DM are common in $\Lambda$CDM simulations therefore, 1E0657-56 should not be considered a challenge.

In this paper, we take a similar approach to that of \citet{Lee10}, and examine large $\Lambda$CDM $N$-body simulations to see how common these high relative velocities are among massive DM halos.  
One of the things that the earlier works have not performed is an examination of resolution and box size effect. Therefore we first conduct a study to determine the effects of increasing resolution or varying box sizes on the parameters of interest.
We then examine our largest simulation in search for a pair matching the initial conditions required by \citet{Mast08} to reproduce the observed properties of 1E0657-56.

The rest of the paper is organized as follows:
Section\,\ref{sec:simulation} discusses simulation parameters,  
Section\,\ref{sec:analysis} shows the simulation results and examines the distribution of parameters relevant to this study, such as halo masses, pairwise velocity, and pair separation distances.  
Section\,\ref{sec:earlyz} examines the simulation results at earlier redshifts of $z$=0.296 \& $z$=0.489, 
and how they relate to the bullet system. 
Finally, Section\,\ref{sec:discussion} contains concluding remarks and discussion of future prospects.


\section{Simulations}
\label{sec:simulation}

For our simulations we use the {\small GADGET-3} code \citep[originally described in][]{Springel05} which simulates large $N$-body problems by means of calculating gravitational interactions with a hierarchical multipole expansion.
It uses a particle-mesh method \citep{Hockney81,Klypin83,White83} For long-range forces and a tree method \citep{BarnesHut86} for short-range forces.

Cosmological parameters consistent with the cosmological constraints from the Wilkinson Microwave Anisotropy Probe {\small (WMAP)} data and an \citet{EandHu98} transfer function were employed when creating the initial conditions for each simulation with random Gaussian phases: ($\Om$, $\Ol$, $H_{0}$, $\sigma_{8}$, $n_{s}$)$=$(0.26, 0.74, 72, 0.8, 1.0) \citep{WMAP5,WMAP7}.  
We note that we used a value of $n_s$=1.0 although the best-fit value from the WMAP data is $n_s$=0.96.  Two additional simulations were ran with $n_s$=0.96 and no differences in their high mass halo pairs were found from the $n_s$=1 simulations, because the tilt in the primordial power spectrum mostly changes the small scale structures. 
All simulations contain only collision-less dark matter particles that interact solely through gravity.

Several simulations with varying particle counts and box sizes were ran from $z$=100 to $z$=0. 
The list of simulations along with other parameters
can be found in Table \ref{table:sim}.  
Starting with the L250N125 run, the box size and particle count were simultaneously increased (from $L_{\rm box}=250\himpc$ to 2016$\himpc$, and from $N=125^3$ to 1008$^3$ particles) 
in order to maintain the same mass resolution and gravitational softening length up until the L2016N1008 run.  The second set of simulations were ran to examine the resolution effect.  We started with the original L250N125 simulation and increased the particle count and decreased the gravitational softening length while keeping the box size the same, up to the L250N500 run.  

\begin{figure}
\includegraphics[scale=0.43]{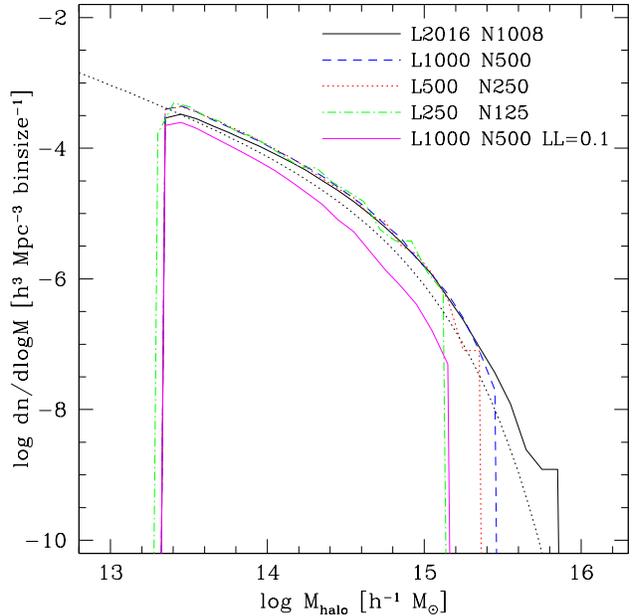}
\caption{DM halo mass function at $z$=0.  This figure shows the box size effect; how increasing the simulation box size allows for a larger number of high mass halos.  The abscissa uses a bin size of $\Delta \log \Mhalo$=0.1.  The black dotted line is the ST mass function using the \citet{EandHu98} transfer function.  The solid magenta line is from the 1GpcN500 simulation grouped with a linking length parameter of $b$=0.1 instead of 0.2.}
\label{fig1}
\end{figure}


\section{Data Analysis \& Results}
\label{sec:analysis}

\subsection{Halo Mass Function}
\label{sec:massfunc}

DM particles were grouped using a simplified version of the parallel friends-of-friends (FOF) group finder {\small{SUBFIND}} \citep{Springel01}.
The code groups the particles into DM halos if they lie within a specified linking length (FOF LL).  This linking length is a fraction of the initial mean inter-particle separation, for which we adopt a standard value of $b$=0.2.
In order to be considered a halo it must also contain at least 32 particles.  

\begin{figure}
\includegraphics[scale=0.43]{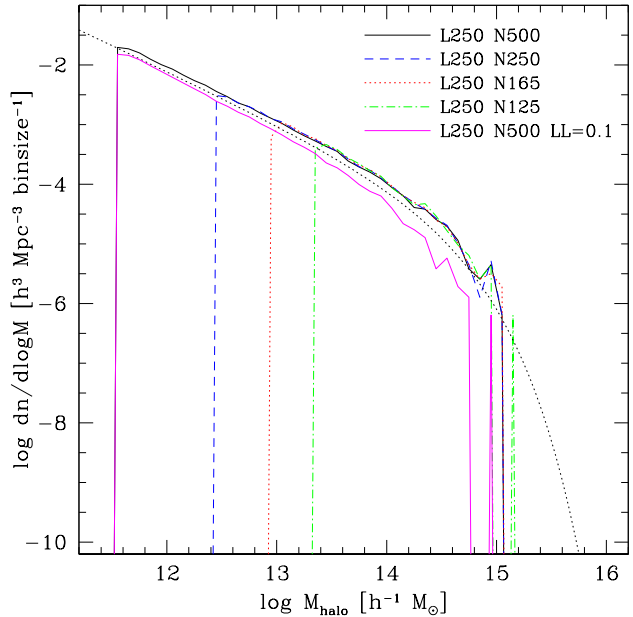}
\caption{DM halo mass function at $z$=0.  This figure shows the resolution effect; how increasing the resolution of a simulation allows for a greater number of small mass halos.  
The abscissa uses a bin size of $\Delta \log \Mhalo$=0.1.  The black dotted line is the ST mass function using the \citet{EandHu98} transfer function.  The solid magenta line is from the 250MpcN500 run, using a linking length of $b$=0.1 instead of 0.2.}
\label{fig2}
\end{figure}

Figures \ref{fig1} and \ref{fig2} show DM halo mass functions in our simulations. 
Both figures include the \citet{ST99} mass function (ST) plotted as a black dotted line. 
Recent work by \citet{More11} found that the commonly used value of $b$=0.2 selects a significantly larger local overdensity ($\delta_{\rm{FOF}}$) than previously thought.
Normally it is assumed that $b$=0.2 results in $\delta_{\rm{FOF}}\approx 60$ (corresponding to the enclosed overdensity of $\delta \sim 180$), but their study finds that it results in $\delta_{\rm{FOF}}\approx 80.61$ which is a $\sim$35\% increase.
We find that our mass function is slightly higher than the ST mass function on all mass scales. 
By regrouping the L1000N500 sim using $b$=0.1 we under-predict the number density on all mass scales, as shown by the solid magenta line in Figures~\ref{fig1} \& \ref{fig2}. 
Changes in $b$ certainly have a significant impact on the halo mass function. 

Figure~\ref{fig1} shows that the number of high mass halos increases by increasing the box size from $250\ h^{-1}$Mpc to $2016\ h^{-1}$Mpc while maintaining the same resolution. 
The lowest mass halo in all simulations shown in Figure~\ref{fig1} is $M_{\rm{halomin}}$=$1.84\times10^{13} h^{-1} \Msun$.  
The run with the largest box size (L2016) shows a slight shortage in the number of low mass halos around $\Mhalo$$\simeq 10^{13.24}-10^{14.20}h^{-1} M_{\odot}$ when compared to the other three runs with smaller box sizes.
The most likely explanation for this shortage is that the lower mass halos were absorbed into higher mass halos.

Higher resolution runs can resolve larger number of low mass halos as seen in Figure~\ref{fig2}.  The least massive halo for the highest resolution simulation (L250N500) has $M_{\rm{halomin}}$=$2.87\times10^{11}h^{-1} \Msun$, which is roughly two orders of magnitude lower than the lowest mass halos found in Figure~\ref{fig1}. 

While searching for a bullet-like pair of halos with masses on the order of $M_{\rm{bullet}}$ \& $M_{\rm{parent}}$, Figures \ref{fig1} \& \ref{fig2} indicate that it is possible to form such massive halos in box sizes as small as $250\ h^{-1}$Mpc at $z$=0 but there will be a low number of them.  

\begin{figure}
\includegraphics[scale=0.43]{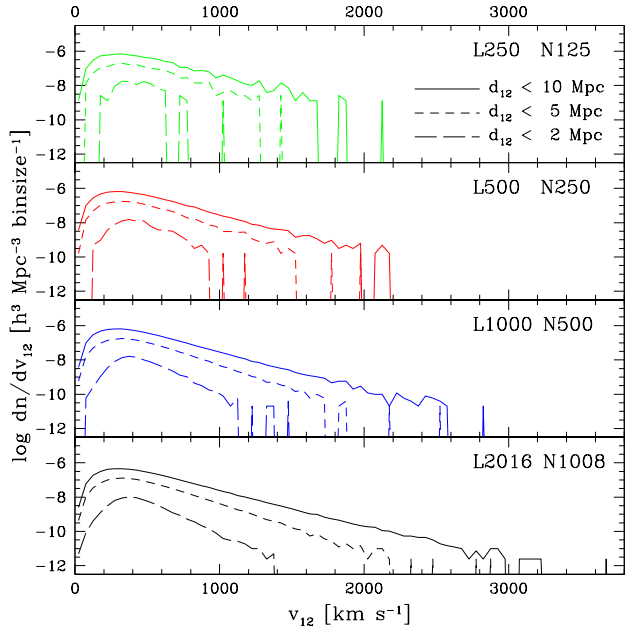}
\caption{Pairwise velocity function at $z=0$, demonstrating the box size effect.  Each panel contains three lines representing pair separation distances of $d_{12}=2$, 5, \& 10 Mpc.  Increasing the box size allows a higher $v_{12}$ for pairs within $d_{12}<10$\,Mpc, while pairs within $d_{12}<5$\,Mpc only see a minor increase.  Pairs residing within $d_{12}<2$\,Mpc see the smallest increase in $v_{12}$ as the box size increases. 
}
\label{fig3}
\end{figure}

\subsection{Pairwise Velocity Function}
\label{sec:v12}

\begin{figure}
\includegraphics[scale=0.43]{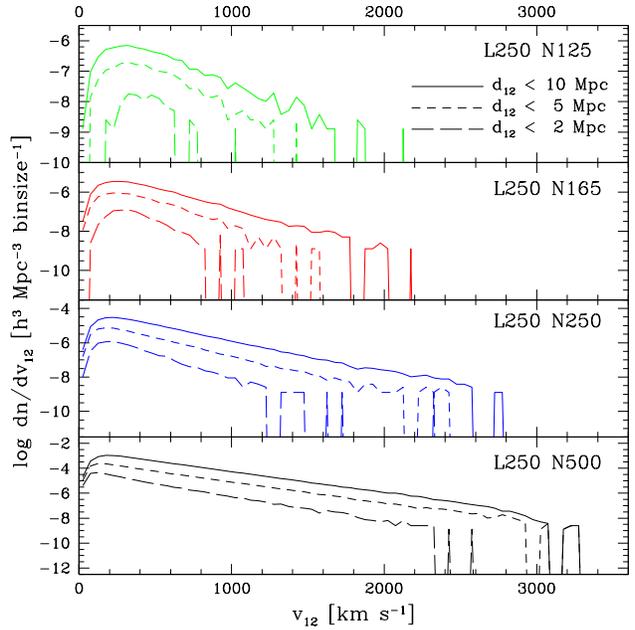}
\caption{Pairwise velocity function at $z$=0, demonstrating the resolution effect.  Each panel contains three lines representing pair separation distances of $d_{12} =2$, 5, \& 10\,Mpc.
Each subsequent increase in resolution allows for smaller structures to be resolved, leading to the increase in $v_{12}$ at all separation distances.
}
\label{fig4}
\end{figure}

In this section, we present the results on the pairwise velocity ($v_{12} = |\vec{v}_1 - \vec{v}_2|$) function, i.e., the number of halo pairs within a velocity bin per unit volume ($dn/dv_{12}$). 
Figures~\ref{fig3} \& \ref{fig4} show $dn/dv_{12}$ with four panels for different simulation runs, each panel containing three lines for halo pairs with a separation distance of less than $d_{12}=2, 5,\ \& \ 10 \ $Mpc.

Figure~\ref{fig3} shows that increasing the box size with a fixed resolution allows for a greater number of high $v_{12}$ pairs, but with greater separation distances.
Doubling the box size from L250 to L500 yields only a small increase in high $v_{12}$ pairs.  Doubling it again to L1000 gives us a considerable jump in high $v_{12}$ pairs with separation distances of 5$<d_{12}<$10 Mpc, while the 2$<d_{12}<$5 Mpc range only sees a moderate increase.  Doubling the box size one final time to L2016, we again only see a moderate increase in $v_{12}$ similar to going from the L250 to L500 sim.  The number of close halo pairs with $d_{12}<$2 Mpc seem to remain fairly constant with relatively low $v_{12}$ throughout changes in the box size.  This implies that increasing the box size does not increase $v_{12}$ for pairs within 2 Mpc of one another.

By increasing the resolution, the number of halo pairs with high $v_{12}$ increases (Figure~\ref{fig4}), but unlike the case of enlarging the box, this does not necessarily come at the cost of increased separation distances.  Each increase in resolution gives us a larger number of low and high $v_{12}$ pairs on all distance scales.
When compared to Figure~\ref{fig3}, the simulations shown in Figure~\ref{fig4} are better at resolving smaller structures and length scales, leading to larger values of $v_{12}$.
Unfortunately this data does not give us any information on the mass of the halos pairs, so increasing the resolution in order to increase the number of close high $v_{12}$ pairs may not be beneficial when searching for high mass pairs such as 1E0657-56.

\subsection{Relative Halo Velocity \& Halo Mass}

\begin{figure}
\includegraphics[scale=0.43]{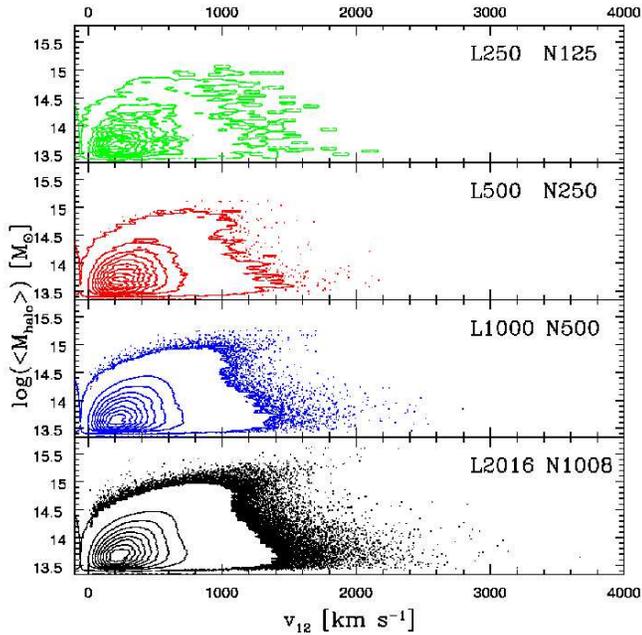}
\caption{Pairwise velocity vs. average mass of DM halo pairs at $z$=0.  Here we show the box size effect; increasing the simulation box size increases the number of low-mass, high-$v_{12}$ pairs more than the high-mass, high-$v_{12}$ pairs.  Each increase in the box size and particle count yields better statistics, broadening the distribution of $v_{12}$.}
\label{fig5}
\end{figure}

It is useful to study the effects of different box sizes and resolutions on the average mass of a halo pair vs. $v_{12}$.
Figure~\ref{fig5} shows how increasing the box size with a constant resolution increases the number of low-mass, high $v_{12}$ halo pairs, along with increasing the number of high-mass, high-$v_{12}$ pairs to a lesser degree.  
As the box size increases, we are allowing for a greater number of rare high $v_{12}$ halo pairs which probe the tail of the distribution.

\begin{figure}
\includegraphics[scale=0.43]{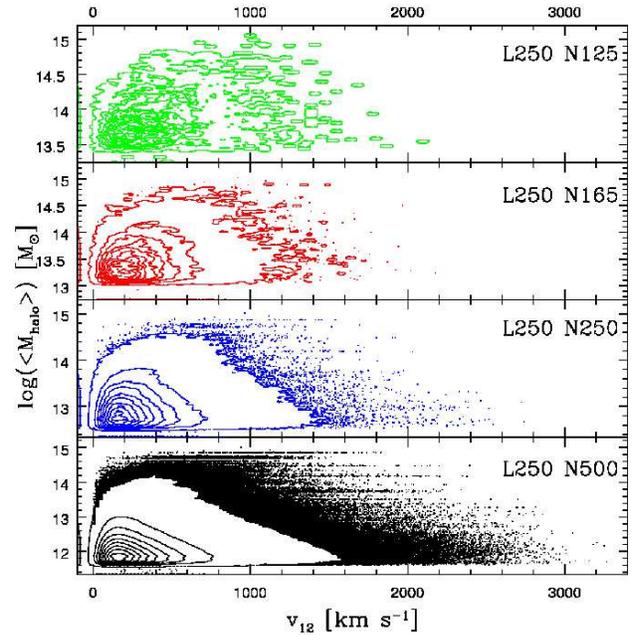}
\caption{Pairwise velocity vs. average DM halo pair mass at $z$=0.  This illustrates the resolution effect; how increasing the resolution probes lower mass halo pairs.  There is a slight increase in high-mass, high-$v_{12}$ pairs, but the majority of the increase is in the low mass halos.  As the particle count increases we can resolve smaller structures with higher $v_{12}$.}
\label{fig6}
\end{figure}

Figure~\ref{fig6} shows that an increase in the resolution results in a larger number of low-mass, high-$v_{12}$ pairs, and a less substantial increase in the number of high-mass, high-$v_{12}$ pairs.
Increasing the box size yields high $v_{12}$ pairs with increasing mass, while increasing the resolution yields a larger number of high $v_{12}$ pairs at the maximum halo mass allowed by the box.

\begin{figure}
\includegraphics[scale=0.43]{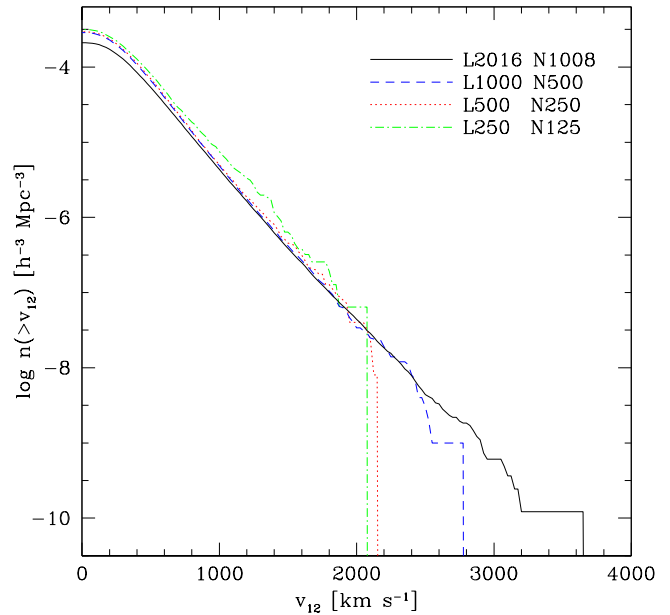}
\caption{Cumulative $v_{12}$ function of DM halos at $z$=0.  This figure shows how increasing the box size increases the number of high-$v_{12}$ pairs, extending the tail of the distribution.   
}
\label{fig7}
\end{figure}

\subsection{Cumulative $v_{12}$ Function}

\begin{figure}
\includegraphics[scale=0.43]{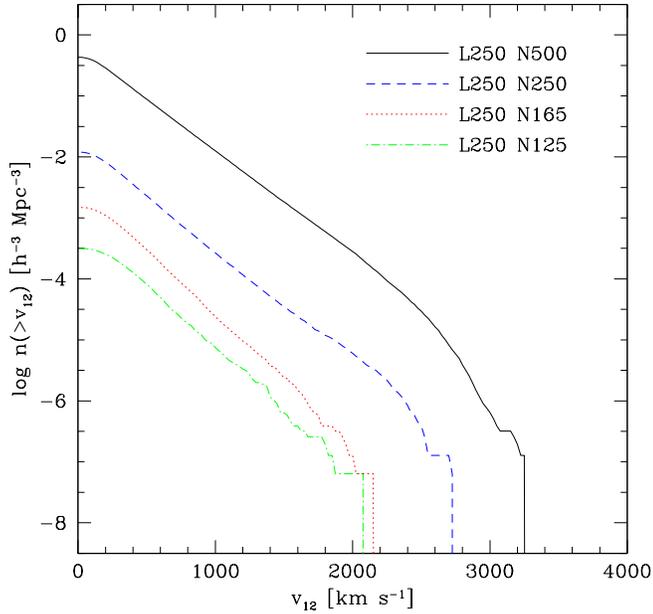}
\caption{Cumulative $v_{12}$ function of DM halos at $z$=0.  This figure shows the resolution effect. As the resolution increases, the normalization of the distribution increases due to a larger number of lower mass halos with higher velocities. 
}
\label{fig8}
\end{figure}

To examine how the box size and resolution affects the actual number of high $v_{12}$ halo pairs, we plot the cumulative $v_{12}$ distribution function as shown in Figures~\ref{fig7} \& \ref{fig8}.  
Changing the box size (Figure \ref{fig7}) extends the curve to higher $v_{12}$. The larger box and particle count result in better statistics, which allows us to probe the high velocity tail of the $v_{12}$ distribution as mentioned in the previous section.

By increasing the resolution alone (Figure~\ref{fig8}), we see that the normalization of the cumulative $v_{12}$ distribution function becomes higher due to larger number of lower mass halos. 
These figures suggest that by increasing the box size and/or resolution one would be able to produce a halo pair with a greater $v_{12}$, however, 
as we saw earlier in Figures \ref{fig5} \& \ref{fig6}, the majority of high-$v_{12}$ pairs have lower average masses than 1E0657-56.

\section{Results at Earlier Redshifts}
\label{sec:earlyz}

To be fully consistent with the observations of 1E0657-56, comparing our simulations at the same redshift as 1E0657-56 would be ideal.  Up until this point, we have examined only simulation data at $z$=0, yet 1E0657-56 is observed at $z$=0.296.  This difference in time of $\sim$3.31 billion years can have a considerable impact on the velocities, sizes, and separation distances of the DM halos contained in the simulation.  
Another problem arises when we consider how we group the DM particles.  
At $z$=0.296 the separation between the two halos of 1E0657-56 is $d_{12}\simeq 0.72$\,Mpc, which is larger than the linking lengths listed in Table~\ref{table:sim} (0.1-0.4 Mpc) for each of our simulations.  
At first glance it may appear that we could identify each halo independently within our sims, but when one considers their large masses, we find that this is not the case.
The virial radius of each halo is found to be 1.42 \& 3.06 Mpc for the `bullet' ($M_{\rm{bullet}} \simeq 1.5 \times 10^{14} \Msun$) and its `parent' ($M_{\rm{parent}} \simeq 1.5 \times 10^{15} \Msun$), respectively.  When two halos of this size are separated by $\simeq$0.72\,Mpc, they will easily overlap, resulting in the FOF group finder identifying them as a single halo at the observed redshift of $z$=0.296. 
If we assume the separation distance of $5$\,Mpc and infall velocity of $3000$\,km\,s$^{-1}$ as required by \citet{Mast08} to reproduce the observed quantities of 1E0657-56, then a halo pair in this initial configuration should be found at $z$=0.489.

\subsection{Peculiar Velocities}
\label{sec:peculiar}

Before we examine the simulation at $z$=0.489, we first compare our simulations to the prediction of linear theory for further validation.
Linear theory predicts that for an Einstein de-Sitter (EdS) universe the growing mode of the peculiar velocity field grows as $t^{1/3}$.  The peculiar velocity of each mode in a non-EdS universe is given by \citep{PeeblesLSS}
\begin{equation}
v_{pec}=\frac{H(z) \, a^2}{4\pi}\frac{dD}{da},
\label{eq:peebles}
\end{equation}
where $H(z)$=$H_0E(z)$ is the Hubble parameter, $a$ is the scale factor, $D$ is the growth factor for linear perturbations, and $E(z)$=$[\Omega_{m,0}(1+z)^3 +(1-\Omega_{k,0}-\Omega_{m,0})(1+z)^2+\Omega_{\Lambda,0}]^{1/2}$.

\begin{figure}
\includegraphics[scale=0.43]{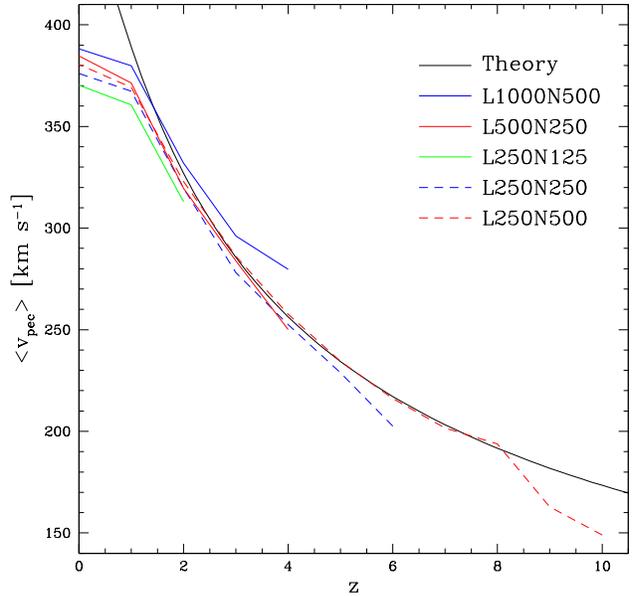}
\caption{The average halo peculiar velocity of five simulations used in this study, compared with the normalized prediction of linear theory described by Eq.~(\ref{eq:peebles}).
The data agrees with theory at $z>1$, but the velocities begin to level off at $z<1$, which is likely due to the virialization of the halos.}
\label{fig9}
\end{figure}

The peculiar velocity of each halo in five of our runs was calculated and averaged up to $z$=10, then compared against the normalized theory curve in Figure \ref{fig9}.  
Our simulations agree well between $z$=6 to $z$=1.0, but start to deviate from the linear theory curve at $z<1.0$, which is likely due to their virialization.

\subsection{Pairwise Velocity: Linear Theory}
\label{sec:pairwise}

\citet{Jusz99} proposed a simple closed-form expression relating the mean relative velocity of pairs of galaxies at a fixed separation to the two-point correlation function of mass density fluctuations:
\begin{equation}
-\frac{v_{12}}{Hr}\approx \frac{2}{3}f \Bar{\Bar{\xi}} \left[ 1+\alpha \bar{\bar{\xi}} \right],
\label{eq:jusz}
\end{equation}
where $H$ is the Hubble parameter, $r=ax$ is the proper separation, $f\equiv d\ln D/d \ln a$, 
$\alpha \approx 1.2-0.65\gamma$, $\gamma$ is the logarithmic slope of the correlation function,
$\Bar{\Bar{\xi}} = \bar{\xi}/\left[1+\xi \right]$, $\bar{\xi} = 3x^{-3}\int^{x}_{0} \xi y^2 dy$, 
and $\xi$ is the two point correlation function.
At $z$=0, the value of $f$ is $\simeq$0.5, and then it asymptotes to unity at $z$$\gtrsim$8.  

\begin{figure}
\includegraphics[scale=0.43]{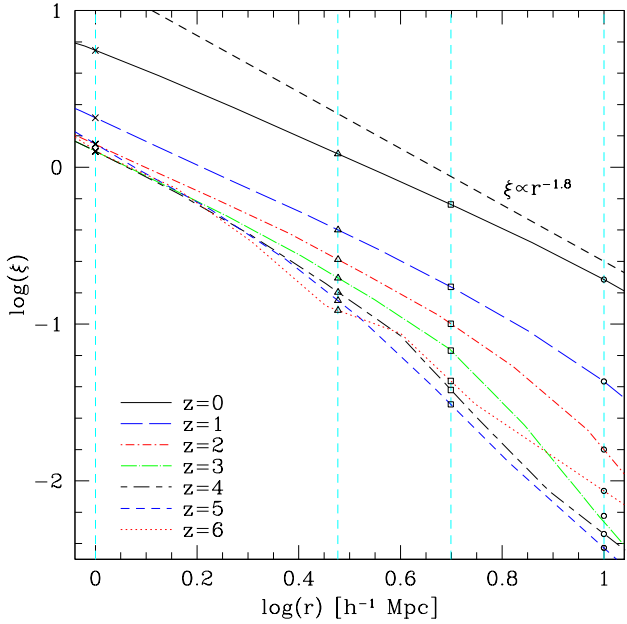}
\caption{Auto-correlation function of DM halos in the L250N500 run at $z$=0$-$6.  The vertical cyan dashed lines indicate $r$=1, 3, 5, \& 10\,Mpc, where we measure the evolution of $\xi$ as a function of redshift.  Symbols lying along these dashed lines represent the $\xi$-values used in Eq.~(\ref{eq:jusz}) for producing the dashed lines in Figure \ref{fig11}.  For comparison, we also show the dashed black line with a slope of $\xi \propto r^{-1.8}$ --- the result consistent with the $z$=0 SDSS galaxies \citep{Zehavi10}.}
\label{fig10}
\end{figure}

To obtain theoretical results based on Eq.~(\ref{eq:jusz}) that can be compared with our simulations, we calculate $\xi$ by correlating the center-of-mass positions of halos with a random data set and use the \citet{Landy93} estimator
\begin{equation}
\xi(r)_{\rm halo}=\frac{DD-2DR+RR}{RR},
\label{eq:correlation}
\end{equation}  
where DD, DR, \& RR represents halo pair counts for Data-Data, Data-Random, \& Random-Random data sets at a given value of $r$.
The result of $\xi(r)_{\rm halo}$ for the L250N500 sim is plotted in Figure~\ref{fig10}.  
Higher values of $\xi_{\rm halo}$ correspond to a larger probability that another halo lies at a separation of $r$.  The value of $\xi_{\rm halo}$ decreases with increasing $r$, implying that halos tend to cluster more on smaller scales.  The value of $\xi_{\rm halo}$ also decreases with increasing redshift, meaning halos are less clustered in the earlier universe.

\begin{figure}
\includegraphics[scale=0.43]{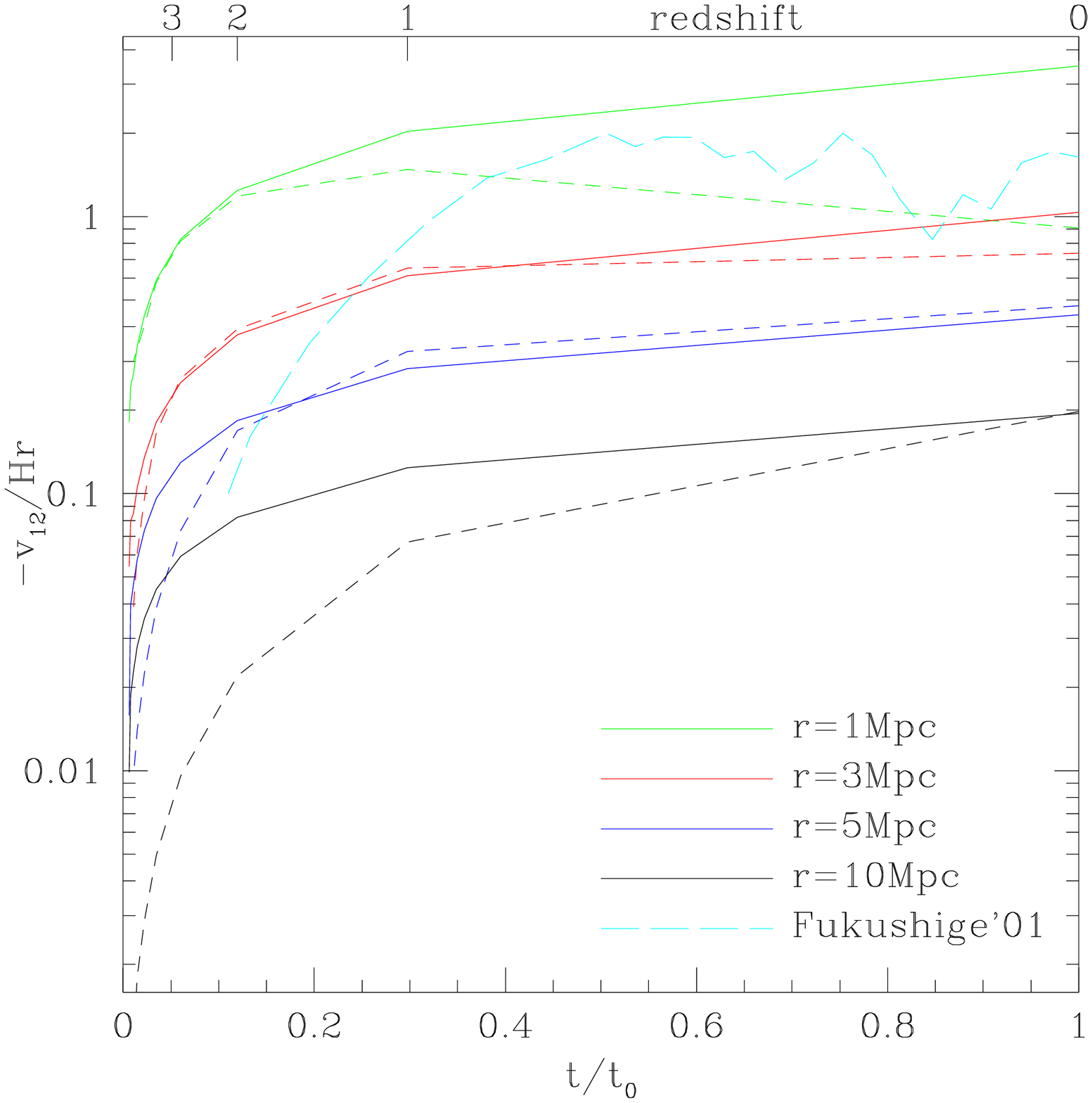}
\caption{{\it Solid lines:} Average pairwise halo velocities $\avg{v_{12}}$ from the 250MpcN500 run residing in physical shells of 1Mpc thickness with the indicated radii.  {\it Dashed lines:} Theoretical $\avg{v_{12}}$ curves given by Equation (\ref{eq:jusz}) using the $\xi$ values from Figure \ref{fig10} at each corresponding radius.  The dashed cyan line represents data from \citet{Fuku01} at a separation distance of r=1.52Mpc.
When these curves reside below unity the Hubble flow is greater than their pairwise velocities, at unity their physical separations remain constant, and above unity their pairwise velocities are greater than the Hubble flow.
}
\label{fig11}
\end{figure}

To compare our simulation with Eq.~(\ref{eq:jusz}), 
we calculated the average pairwise halo velocities $\avg{v_{12}}$ for pairs residing within physical shells of 1\,Mpc thickness ($\pm$0.5\,Mpc) around $r$=1, 3, 5, \& 10 Mpc for the L250N500 run.
The results are shown in Figure~\ref{fig11}, where the solid curves represent simulation data, the dashed curves correspond to the theoretical predictions of Eq.~(\ref{eq:jusz}) using $\xi$-values from Figure~\ref{fig10}, and the different colors distinguish between different values of $r$.
\citet{Jusz99} did not consider the effect of galaxy bias relative to dark matter, and without any correction, we find that $\avg{v_{12}}$ of halos in our simulation are somewhat higher than those predicted by Eq.~(\ref{eq:jusz}). Therefore we invoke an ad hoc correction factor of $\times$1.5 to account for this effect, and the dashed lines in Figure~\ref{fig11} include this multiplication factor in the right-hand-size of Eq.~(\ref{eq:jusz}).  
After this correction, our simulation agrees with Eq.~(\ref{eq:jusz}) very well for $r=3$ \& 5\,Mpc, but there is some deviation from theory for the $r=1$ \& 10\,Mpc results. 
The shape of the theory curve is determined by the competition between increasing $H(z)$, decreasing $\xi$, and increasing $f$ with increasing redshift.

\citet{Fuku01} examined the validity and limitations of the stable condition ($-v_{12}/Hr=1$), which states that the mean physical separation $r$ of galaxy pairs is constant on small scales.  They found a significant time variation in the mean pairwise peculiar velocities and argued that this behavior was not due to a numerical artifact, but a natural consequence of the continuous merging process.  This irregular oscillatory behavior could be reduced by averaging over cosmological volumes larger than 200 Mpc$^3$, resulting in a more accurate estimate of the mean pairwise velocity.
Our data is also consistent with \citet{Fuku01} (dashed cyan line in Figure~\ref{fig11}) in that the oscillatory behavior is suppressed due to our cosmological volume being greater than 200 Mpc$^3$, and their result for $r$=1.52\,Mpc lies between our $r$=1 \& 3\,Mpc curves.

\subsection{In Search of the `Bullet'}

\begin{figure}
\includegraphics[scale=0.43]{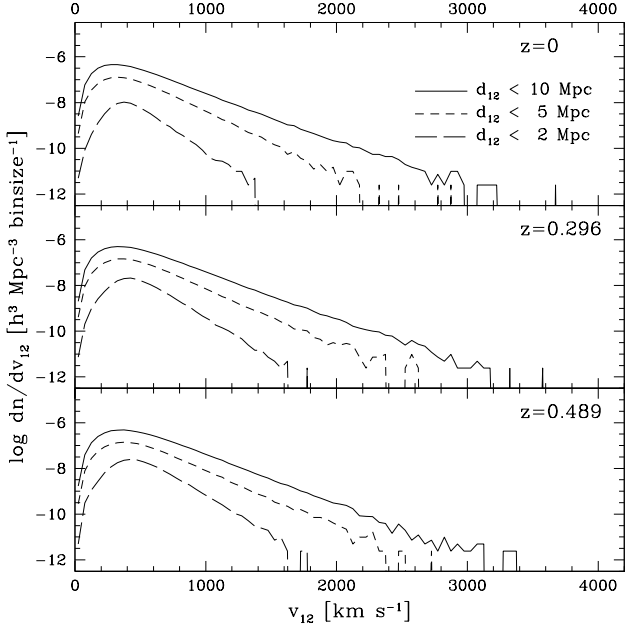}
\caption{
Pairwise velocity function for the L2016N1008 run at $z$=0.0, 0.296, \& 0.489. 
There is a slight increase in the number of pairs at the highest end of the $v_{12}$ distribution as the redshift increases. 
}
\label{fig12}
\end{figure}

Hereafter we will only be examining our largest simulation {\small{(L2016N1008)}} at redshifts of $z$=0.0, 0.296, and 0.489.
In Figure~\ref{fig12}, we show the redshift evolution of the pairwise velocity function ($dn/dv_{12}$) from $z$=0 to $z$=0.489.
Qualitatively this plot changes very little with redshift, except that there is a slight increase in the number of pairs at the highest end of the $v_{12}$ distribution. 
Pairs within separation distances of $d_{12}<2$\,Mpc have maximum $v_{12}$ on the order of $\simeq$1800 km s$^{-1}$ at $z$=0.296 and $z$=0.489.
For pairs with greater $d_{12}$, the maximum $v_{12}$ reaches as high as $\simeq$3300 km s$^{-1}$.

\begin{figure}
\includegraphics[scale=0.43]{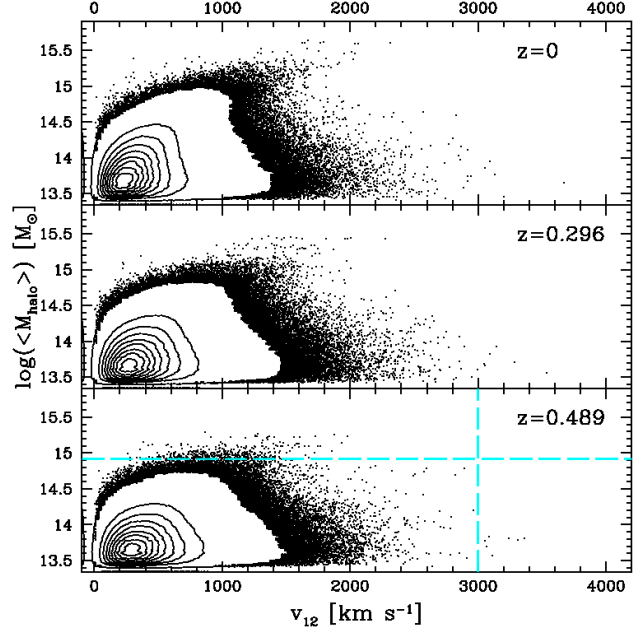}
\caption{
Average mass of halo pairs vs. their pairwise velocity for the L2016N1008 run at $z$=0.0, 0.296, \& 0.489.  In the bottom panel ($z$=0.489) the horizontal dashed line represents an average pair mass of $8.25\times 10^{14} \Msun$ for 1E0657-56, and the vertical dashed line represents a pairwise velocity of 3000\,km\,s$^{-1}$ suggested by \citet{Mast08}.
}
\label{fig13}
\end{figure}

In Figure~\ref{fig13}, we show the redshift evolution of the average halo mass vs. their pairwise velocity.  
One can see the effect of halo mergers, and the number of high-mass halo pairs with $\avg{\Mhalo} > 10^{15}\Msun$ are increasing from $z$=0.489 to $z$=0. 
The cyan dashed lines in the $z$=0.489 panel illustrate the average pair mass of 1E0657-56 ($8.25 \times 10^{14} \Msun$) and initial pairwise velocity of $v_{12}\approx 3000$\,km\,s$^{-1}$ required by \citet{Mast08}.
Two pairs are found in our simulation near the region of interest, but their masses and velocities are still too low.

\subsubsection{Candidate Halo Pairs}

Table~\ref{table:masspairs} lists the five halo pairs with highest $\avg{\Mhalo}$ for $z$=0, $z$=0.296, and $z$=0.489.  
A simulation of this size (comoving 2\,$h^{-1}$Gpc) produces many halo pairs massive enough to match that of 1E0657-56 at the examined redshifts.
While the separation distances of these pairs may be in the range we are interested in, the pairwise velocities are too low to match the required $v_{12}$=3000\,km\,s$^{-1}$ by \citet{Mast08}. 

Table~\ref{table:v12pairs} lists the five halo pairs with the highest $v_{12}$ at the three examined redshifts.  
All halo pairs in this table match or exceed the required $v_{12}$ of 3000\,km\,s$^{-1}$, but they miss the mark when it comes to the other observables of 1E0657-56.  All of the halos in this table have masses one or two orders of magnitude lower than $M_{\rm{bullet}}$ \& $M_{\rm{parent}}$.
The mass ratios are also a bit high; the lowest being $\sim$0.3 at $z$=0.489 compared to 0.1 for 1E0657-56 at $z$=0.296.
None of the collision angles are head on, yet most are highly inclined.
Lastly the separation distance of each pair at $z$=0.489 is somewhat large; 
\citet{Mast08} set their initial separation at proper 5\,Mpc while each pair in this table is separated by $>$7.5\,Mpc.

\subsubsection{Simulation Requirements to Produce the `Bullet'}
\label{sec:cumfit}

\begin{figure*}
\includegraphics[scale=0.6]{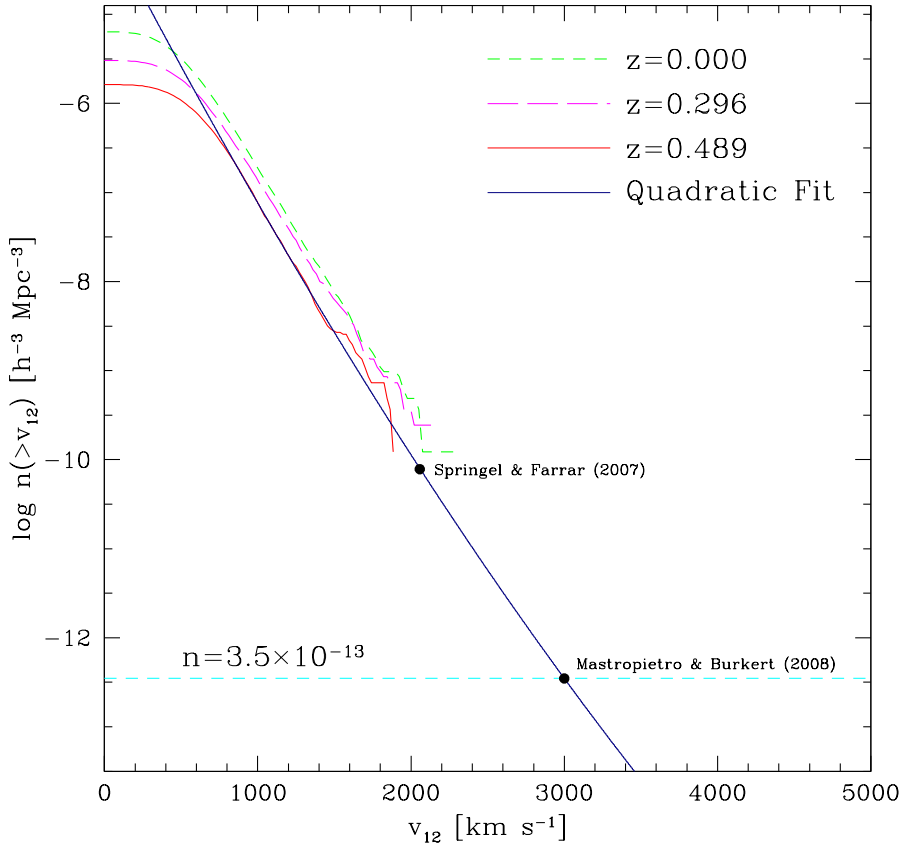}
\caption{Comoving number density of halo pairs in the N2016N1008 run with masses above 10$^{14} \Msun$ at $z$=0, 0.296,\& 0.489.
We also over-plot a quadratic fit described in the text for $z$=0.489. 
The horizontal dashed line illustrates the number density of halos with $v_{12}=3000$\,km\,s$^{-1}$ corresponding to a box size of (4.48\,$h^{-1}$\,Gpc)$^3$ and $2240^3$ DM particles.
The black filled circles represent the $v_{12}$ values listed in Table~\ref{table:cumfit}.
}
\label{fig14}
\end{figure*}

In Figures~\ref{fig7} \& \ref{fig8}, we examined the cumulative $v_{12}$ distribution, however these figures included a large number of low-mass halos which are of little interest to this study.
Therefore in Figure~\ref{fig14}, we restrict the halo sample to those with masses greater than $10^{14} \Msun$ at $z$=0, $z$=0.296, \& $z$=0.489.  
With increasing redshift we see a decrease in the total number density of halo pairs above $10^{14} \Msun$.  

Assuming that the trend of the cumulative $v_{12}$ function would continue to higher velocities with increasing box size (as was the case for $z$=0 shown in Figure~\ref{fig7}), we can fit a line to the $z$=0.489 curve and estimate the box size and particle count required to produce at least one halo pair with a specified $v_{12}$.  
A quadratic of the form $y=y_0+ax+bx^2$ was fit to the $z$=0.489 curve between the values of $v_{12}$=800$-$1500\,km\,s$^{-1}$, and we obtain the best fit values of $y_0$=$-3.97$, $a$=$-3.31\times10^{-3}$, \& $b$=$1.59\times10^{-7}$. 
Based on this fit, we estimate the minimum box sizes and particle counts (for the same resolution as the L2016N1008 run) required to produce at least one halo with the initial velocities given by 
\citet{Mast08}, and \citet{Springel07}.
The result is listed in Table~\ref{table:cumfit}. 

Our result suggests that we would need a simulation box size of $(4.48 h^{-1}\rm{Gpc})^3$ \& $2240^3$ DM particles in order to produce at least one halo pair with an average mass greater than 10$^{14} \Msun$ and $v_{12}>3000$\,km\,s$^{-1}$ at $z$=0.489.
The exact values of the required box size and particle count is somewhat sensitive to the range of $v_{12}$ used for the fit, therefore the values listed in 
Table~\ref{table:cumfit} should be taken as a rough estimate. 
The required simulations are so large and they would take significant computational resources which is currently not feasible for us.

\subsubsection{Probability of Finding the `Bullet'}
\label{sec:pdf}

We also examine the probability distribution function (PDF) of $v_{12}$
for halos with $\avg{\Mhalo}>10^{14} \Msun$. 
We perform a least square fit to the data using a skewed normal distribution \citep{SkewNormal}, and calculate the probability of finding a halo pair with $v_{12}>3000$\,km\,s$^{-1}$ at $z$=0.489.

\begin{figure}
\includegraphics[scale=0.43]{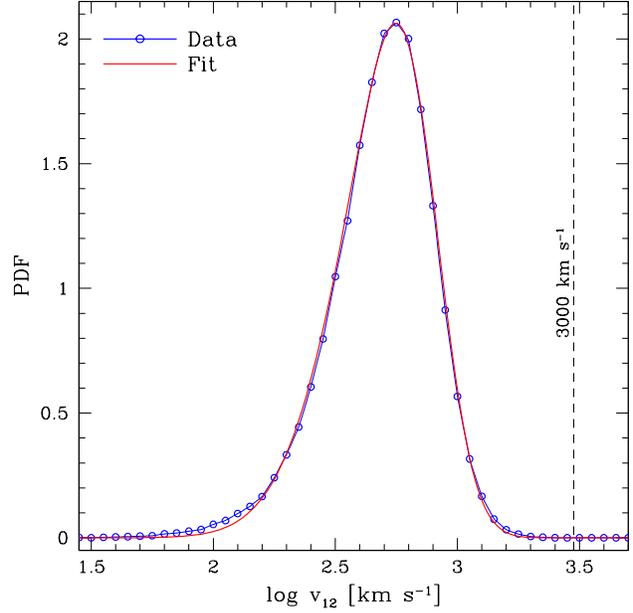}
\caption{Pairwise velocity probability distribution function for halo pairs with masses above 10$^{14} \Msun$ in our L2016N1008 run.  The blue circles represent $v_{12}$ binned PDF data, the blue curve is the linearly interpolated values, and the red curve is the best-fit skew normal distribution \citep{SkewNormal}.  Integrating the fit from $v_{12}=3000$\,km\,s$^{-1}$ to infinity gives $P(>3000$\,km\,s$^{-1})=2.8\times10^{-8}$.  This very low probability suggests that it is very difficult to produce a halo pair with high mass and high-$v_{12}$ as the observed 1E0657-56. }
\label{fig15}
\end{figure}

In Figure~\ref{fig15}, we show the binned PDF data with blue circles, and the best-fit skew normal distribution as the red curve.  
By integrating the PDF from $v_{12}=3000$\,km\,s$^{-1}$ to infinity, we calculate the probability of finding a halo pair with masses greater than 10$^{14}\,\Msun$ and $v_{12}>3000$\,km\,s$^{-1}$ to be $P(>3000$\,km\,s$^{-1})=2.8 \times 10^{-8}$, 
which is roughly one order of magnitude higher than calculations done by \citet{Lee10} ($P$=3.6$\times10^{-9}$). 
This very low probability corroborates our earlier finding that it is very difficult to produce a massive halo pair with a high $v_{12}$ matching the required initial configuration suggested by \citet{Mast08}.


\section{Conclusions}
\label{sec:discussion}

We performed many $N$-Body cosmological simulations with varying box sizes and resolutions in order to examine how changing these parameters affect the search for high-$v_{12}$ halo pairs comparable to 
the initial conditions required to reproduce the observed properties of the 1E0657-56 system in non-cosmological simulations. 
Using our largest L2016N1008 run, we examined the pairwise velocities, halo masses, and halo separation distances at $z$=0.0, 0.296, \& 0.489. 

We find that the high-$v_{12}$ tail of the distribution extends to a greater velocities as we increase the simulation box size. We also find that the number of high-$v_{12}$ pairs increased as we increase the particle count and resolution with a fixed box size, however, this increase is mostly due to lower mass halos which do not correspond to the characteristics of 1E0657-56.  We find that the redshift evolution effect is not very strong for the $v_{12}$ distribution function. 

As we show in Table~\ref{table:v12pairs}, some of the halo pairs have a high relative velocity similar to 
the initial conditions required to reproduce the observational quantities of 1E0657-56 in non-cosmological simulations,
but they are galaxy group-scale halos (10$^{13}$$-$10$^{14}\Msun$) and much less massive than the observed estimates for 1E0657-56.

We find that, in $N$-body simulations with comoving volumes of less than (2\,$h^{-1}$\,Gpc)$^3$, it is very difficult to reproduce a system that resembles the initial conditions required to reproduce the observational properties of 1E0657-56.  Based on the extrapolation of our cumulative $v_{12}$ function, we find that one needs a simulation with a comoving box size of (4.48\,$h^{-1}$\,Gpc)$^3$ and $2240^3$ DM particles in order to produce at least one pair of halos that resembles the initial conditions suggested by \citet{Mast08}. 
In the future it would be useful to run larger simulations (e.g., with $\sim$5\,Gpc box and $\sim$2500$^3$ particles) to improve the statistics of massive halos. 

From the simulated $v_{12}$ PDF of halos, we calculated the probability of finding a halo pair with $v_{12}\geq 3000$\,km\,s$^{-1}$ and masses $\geq 10^{14} \Msun$ to be $2.76\times10^{-8}$, which is somewhat larger than previous work by \citet{Lee10}.  However, both probabilities are quite small and the difference is negligible. These results 
suggest that a system like 1E0657-56 is currently incompatible with the concordance $\Lam$CDM universe, if its initial condition really requires an initial pairwise velocity of $v_{12}\geq 3000$\,km\,s$^{-1}$. 
As \citet{Lee10} discussed in detail, there seems to be more systems like 1E0657-56 being observed already, which exacerbates the incompatibility in terms of probability. 
One other possibility is that there is something wrong with the referred non-cosmological simulations, and the suggested initial $v_{12}$ must be revised to a lower value.

\section*{Acknowledgements}

RT acknowledges the support provided by the Nevada NASA Space Grant 
Consortium through NASA grant NNX10AJ82H.
This work is also supported in part by the NSF grant AST-0807491, 
National Aeronautics and Space Administration under Grant/Cooperative 
Agreement No. NNX08AE57A issued by the Nevada NASA EPSCoR program, and 
the President's Infrastructure Award from UNLV. 
This research is also supported by the NSF through the TeraGrid resources 
provided by the Texas Advanced Computing Center (TACC)
and the National Institute for Computational Sciences (NICS).

\pagebreak
\begin{onecolumn}

\begin{deluxetable}{ l c c c c c}
\tabletypesize{\footnotesize}
\tablecolumns{6}
\tablewidth{0pc}
\tablecaption{Summary of Simulations}
\tablehead{\multicolumn{1}{c}{Run Name} &  
	\colhead{Box Size} &
	\colhead{Particle Count} &
	\colhead{$M_{\rm{dm}}$} &
	\colhead{$\epsilon$} &
	\colhead{FOF LL} \\
	\colhead{ \ } &
	\colhead{[$h^{-1}$ Mpc]} &
	\colhead{ \ } &
	\colhead{[$h^{-1}$ $\Msun$]} &
	\colhead{[$h^{-1}$ kpc]} &
	\colhead{[$h^{-1}$ kpc]}
}
\startdata
\sidehead{Box Size Effects}
L250\  \ N125 	& 	250 	&	 $125^3$ 	& 	5.74 $\times \ 10^{11}$  &	80 & 400 \\ 
L500\ \  N250 	& 	500 	&	 $250^3$ 	& 	5.74 $\times \ 10^{11}$  & 80  & 400 \\
L1000 N500 		& 	1000 	&	 $500^3$ 	& 	5.74 $\times \ 10^{11}$  & 80	& 400 \\
L2016 N1008 	& 	2016	&	 $1008	^3$& 	5.74 $\times \ 10^{11}$  & 80 & 400 \\
\sidehead{Resolution Effects} 
L250 N125 & 250 & $125^3$ & 5.74 $\times \ 10^{11}$ & 80 & 400 \\ 
L250 N165 & 250 & $165^3$ & 2.50 $\times \ 10^{11}$ & 60.6 & 303 \\
L250 N250 & 250 & $250^3$ & 7.18 $\times \ 10^{10}$ & 40 & 200 \\
L250 N500 & 250 & $500^3$ & 8.97 $\times \ 10^{9}$  &  20 & 100 \\
\enddata
\tablecomments{Summary of simulations employed in this paper.  
$M_{\rm{dm}}$ is the mass of each DM particle,
$\epsilon$ is the comoving gravitational softening length,
and FOF LL is the friends-of-friends linking length.  The top four simulations explore the effects of increasing box size with fixed resolution, while the bottom four explore the effects of increasing resolution with a fixed box size.}
\label{table:sim}
\end{deluxetable}

\pagebreak

\begin{deluxetable}{ccrcccccccc}
\tabletypesize{\footnotesize}
\tablecolumns{9}
\tablewidth{0pc}
\tablecaption{Highest Mass Pairs}
\tablehead{\colhead{Pair} &
	\colhead{$v_{12}$} &
	\colhead{$\theta$} &
	\colhead{$M_1$} &
	\colhead{$M_2$} &
	\colhead{Mass Ratio} &
	\colhead{$d$} &
	\colhead{$\rm{r_{1\ virial}}$} &
	\colhead{$\rm{r_{2\ virial}}$} 
 }
\startdata
\sidehead{$z$=0} 
1	&	1670	&	165	&	5.71E+15	&	5.02E+14	&	0.088	&	8.70	&	5.67	&	2.52	\\
2	&	1792	&	46	&	5.71E+15	&	1.99E+14	&	0.035	&	7.84	&	5.67	&	1.85	\\
3	&	1767	&	75	&	5.71E+15	&	1.01E+14	&	0.018	&	7.63	&	5.67	&	1.48	\\
4	&	1624	&	80	&	5.71E+15	&	7.33E+13	&	0.013	&	7.13	&	5.67	&	1.33	\\
5	&	2316	&	72	&	5.71E+15	&	7.04E+13	&	0.012	&	6.20	&	5.67	&	1.31	\\
\sidehead{$z$=0.296} 
6	&	1360	&	141	&	3.80E+15	&	3.50E+14	&	0.092	& 	9.55	&	4.18	&	1.89	\\
7	&	1533	&	44	&	3.80E+15	&	2.61E+14	&	0.069	& 	6.23	&	4.18	&	1.71	\\
8	&	1486	&	56	&	3.80E+15	&	2.51E+14	&	0.066	& 	10.00	&	4.18	&	1.69	\\
9	&	1425	&	129	&	3.80E+15	&	2.13E+14	&	0.056	& 	6.20	&	4.18	&	1.60	\\
10	&	2007	&	112	&	3.80E+15	&	1.78E+14	&	0.047	& 	5.65	&	4.18	&	1.51	\\
\sidehead{$z$=0.489}
11	&	869	&	91	&	3.28E+15	&	5.59E+14	&	0.170	&	8.78	&	3.70	&	2.05	\\
12	&	1277	&	111	&	2.64E+15	&	1.07E+15	&	0.405	&	8.11	&	3.44	&	2.55	\\
13	&	1875	&	132	&	2.45E+15	&	1.19E+15	&	0.485	&	3.86	&	3.36	&	2.64	\\
14	&	1257	&	108	&	2.45E+15	&	1.08E+15	&	0.440	&	4.83	&	3.36	&	2.55	\\
15	&	1256	&	54	&	3.28E+15	&	1.73E+14	&	0.053	&	6.01	&	3.70	&	1.39	\\
\enddata
\tablecomments{Five halo pairs with the highest average halo mass from the L2016N1008 simulation at $z$=0.0, $z$=0.296 and $z$=0.489.  
The values of $v_{12}$ are given in km\,s$^{-1}$,  collision angles $\theta$ in degrees, masses ($M_1, M_2$) in $\Msun$, pair separation distances ($d_{12}$) and virial radius of each halo in physical Mpc.
Although this simulation can produce massive pairs matching the observed mass of 1E0657-56, these pairs have too low relative velocities, and too large separation distances.  }
\label{table:masspairs}
\end{deluxetable}

\begin{deluxetable}{ccrcccccccc}
\tabletypesize{\footnotesize}
\tablecolumns{9}
\tablewidth{0pc}
\tablecaption{Highest Velocity Pairs}
\tablehead{\colhead{Pair} &
	\colhead{$v_{12}$} &
	\colhead{$\theta$} &
	\colhead{$M_1$} &
	\colhead{$M_2$} &
	\colhead{Mass Ratio} &
	\colhead{$d$} &
	\colhead{$\rm{r_{1\ virial}}$} &
	\colhead{$\rm{r_{2\ virial}}$} &
 }
\startdata
\sidehead{$z$=0}
31	&	3674	&	103	&	3.64E+13	&	2.71E+13	&	0.746	&	8.83	&	1.05	&	0.95	\\
32	&	3199	&	151	&	2.14E+13	&	2.02E+13	&	0.946	&	8.20	&	0.88	&	0.86	\\
33	&	3133	&	134	&	5.83E+13	&	2.60E+13	&	0.446	&	9.09	&	1.23	&	0.94	\\
34	&	3095	&	113	&	8.20E+13	&	4.56E+13	&	0.556	&	9.21	&	1.38	&	1.13	\\
35	&	3053	&	108	&	8.20E+13	&	2.14E+13	&	0.261	&	9.11	&	1.38	&	0.88	\\
\sidehead{$z$=0.296}
36	&	3538	&	143	&	3.35E+13	&	1.96E+13	&	0.586	& 	9.94	&	0.86	&	0.72	\\
37	&	3282	&	125	&	4.96E+13	&	2.37E+13	&	0.477	& 	9.39	&	0.98	&	0.77	\\
38	&	3141	&	155	&	8.60E+13	&	3.41E+13	&	0.396	& 	8.80	&	1.18	&	0.87	\\
39	&	3089	&	170	&	6.93E+13	&	2.77E+13	&	0.400	& 	5.27	&	1.10	&	0.81	\\
40	&	3053	&	153	&	4.16E+13	&	2.48E+13	&	0.597	& 	8.60	&	0.93	&	0.78	\\
\sidehead{$z$=0.489}
41	&	3361	&	128	&	6.75E+13	&	2.60E+13	&	0.385	& 	8.81	&	1.01	&	0.74	\\
42	&	3312	&	148	&	5.66E+13	&	3.18E+13	&	0.561	& 	8.03	&	0.96	&	0.79	\\
43	&	3239	&	102	&	6.75E+13	&	2.37E+13	&	0.350	& 	7.57	&	1.01	&	0.72	\\
44	&	3109	&	146	&	2.94E+13	&	2.37E+13	&	0.804	& 	9.57	&	0.77	&	0.72	\\
45	&	3083	&	103	&	7.56E+13	&	2.37E+13	&	0.313	& 	9.25	&	1.05	&	0.72	\\
\enddata
\tablecomments{
Five halo pairs with highest $v_{12}$ found in the L2016N1008 simulation at $z$=0.0, $z$=0.296 and $z$=0.489.  
The values of $v_{12}$ are given in km s$^{-1}$, collision angles $\theta$ in degrees, masses ($M_1, M_2$) in $\Msun$, pair separation distances ($d_{12}$) and viral radius of each halo in physical Mpc.
None of these high velocity halo pairs are massive enough to match the observations of 1E0657-56.  }
\label{table:v12pairs}
\end{deluxetable}

\pagebreak

\begin{deluxetable}{lccc}
\tabletypesize{\footnotesize}
\tablecolumns{4}
\tablewidth{0pc}
\tablecaption{Simulation Requirements to produce a Bullet}
\tablehead{\colhead{Reference} &
	\colhead{$v_{12}$} &
	\colhead{Box Size} &
	\colhead{Particle Count} \\
	\colhead{ \ } &
	\colhead{[km s$^{-1}$]} &
	\colhead{[$h^{-1}$ Mpc]} &
	\colhead{ \ }
}
\startdata
\citet{Mast08}		&	3000	&	4480	&	$2240^3$ \\
\citet{Springel07}	&	2057	&	2224	&	$1112^3$ \\
\enddata
\tablecomments{
Required box size and particle number needed to produce at least one halo pair with an average mass greater than 10$^{14} \Msun$ and a certain value of $v_{12}$ at $z$=0.489 suggested by each of the authors.  See text in \S~\ref{sec:cumfit} for more details. 
}
\label{table:cumfit}
\end{deluxetable}
\end{onecolumn}

\clearpage

\begin{twocolumn}

\end{twocolumn}
\end{document}